{\bf Conformal Field Theories on the Two-Torus  and  Quotients of $SL2(Z)$ }

\vskip 1.cm  

Antoine D.  Coste*\ 
, Giovanni Felder  
     
\vskip .5cm 
* LPT-CNRS,   UMR 8627,
                   
University building 210 , F 91405 Orsay cedex, France    
         
coste@math.uni-frankfurt.de  

\vskip .5cm 
and FIM  ETH Zentrum , CH-8092  Zurich
              
 Switzerland

\vskip 1.cm   
Abstract         
    
\vskip .5cm 
We present remarkable properties of the groups  $SL2(Z/NZ)$
 which might be useful in detailed studies of some quotients 
  appearing in Conformal Field Theories (CFTs). 

\vskip 1.cm
{\bf Introduction }
\vskip .5cm 

The main object underlying  this study is a finite dimensional representation 
 $ ( \rho , V ) $ of the group 
 $  SL2(Z) $ of matrices $   \pmatrix{  a&b \cr c &d }   $.   
Our (writing) efforts are justified by the high interest for physics 
and mathematics of the space V; indeed there exist an infinity  of such  
representations, a countable subset of explicit examples being provided by 
 integrable representations of infinite dimensional affine algebras,
 (``Kac-Moody''  algebras) and  Virasoro algebra representations 
called ``minimal models''.  Any  of the  rational conformal field theories  
( whenever their classification will be achieved) 
will provide such a representation. 
       
  \vskip .5cm       

In the following paragraphs we look at subgroups or quotients  of 
$SL2(Z) $ expressing elements in terms of words in the two generators  T  and S. 
Physical motivation for this is that $T$ is represented by  a unitary diagonal matrix 
whose eigenvalues are related to the physical dimension of fundamental
 excitations,  whereas $S$  describes the effect of putting a box 
on one of its sides, or permuting the role of ``space'' and  ``time'' 
in some  hamiltonian  description. 

\vskip .5cm 
Since this field of research is very popular, we refer the reader 
to monographies (like the one by J.M. Drouffe and C. Itzykson) and go directly 
into: 
\vskip .5cm 
{\bf Preliminary formulae:}

The best to get acquainted with beauties of matrix groups over rings 
is maybe to let oneself play; therefore 
let us consider products of matrices of the form:  


$$  T^x \ S :=   \pmatrix{1  & x\cr 0   & 1 } 
                                 \pmatrix{0  & -1 \cr 1 & 0}
        \    = \ \pmatrix{x  & -1\cr 1  & 0 }   $$

$$  T^{x_n  }\ S \cdots T^{x_1  }\ S :=  
 \pmatrix{  A_n  &  B_n \cr  C_n &  D_n  } =
 \pmatrix{ x_n\ A_{n-1}- C_{n-1}\  & \  x_n \ B_{n-1} -D_{n-1}\cr 
                A_{n-1}            &       B_{n-1}          }  $$

For arbitrary elements of $SL2(Z)$, we know 
the minimal n required is not bounded, 
 according to the famous Farey's enumeration of rational numbers
  between $0$ and $1$.

For $n=3$ we get: 
$$ T^x \ S    T^{-d }\ S        T^y \ S = 
     \pmatrix{ -(xdy+x+y) & 1+xd \cr -(dy+1) & d }             $$

and also 
$$ T^u \ S    T^{c  }\ S        T^v \  = 
   \pmatrix{uc-1  & ucv-u-v \cr c & cv-1 }                      $$
$$ T^u \ S    T^{-c }\ S^{-1} \   T^v \  = 
   \pmatrix{uc+1  & ucv+u+v \cr c & cv+1 }                     $$

We will always  denote integers with  lower case letters: a,b,c,d  ;
 and by A,B,C,D   their residues modulo  N , a fixed integer equal to 
the order of the matrix $\rho(T) $.   Quite often  
 $ad-bc=1  $  and when $(c,N)=1  $  we will denote by $C^{-1} $ the  
 inverse of  C in $Z/NZ$.   In the case  $(d,n)=1 $ we will 
 denote  its inverse  also by  ${\bf   D^{-1} } $.

\vskip .5cm

In the next paragraphs, we will use some  formulae valid 
for  elements 
$A,B,C,D$   satisfying  
  $ AD-BC\  \equiv \ 1 $ in a commutative ring. 
   (in view of Z/NZ we will denote equality by 
 $\equiv  $, but the formulae are also true  in Z with equality of integers).
\vskip .5cm  
Proposition:  If there exists  $U$ such that $UC\ \equiv A+1    $,  
then  
$$  T^U \ S    T^{C  }\ S        T^{UD-B} \   \equiv
   \pmatrix{A  & B \cr C & D}                                  $$

Proof:  is straightforward from above formula. Similarly we also have:  
\vskip .5cm 
Proposition:  If there exists  $U$ such that $UC\ \equiv A-1 $,  
then  
$$ T^U \ S    T^{-C  }\ S^{-1}        T^{B-DU} \  \equiv  
   \pmatrix{ A  & B \cr C & D}                                 $$
Proposition:  If there exists  $X$ such that $XD\ \equiv B-1 $,  
then  
$$ T^X \ S    T^{-D  }\ S    T^{XC-A} \ S \equiv  
   \pmatrix{ A  & B \cr C & D}                                 $$
Proposition:  If there exists  $X$ such that $XD\ \equiv B+1 $,  
then  
$$ T^X \ S    T^{D  }\ S^{-1}    T^{A-XC} \ S \equiv  
   \pmatrix{ A  & B \cr C & D}                                 $$
Proposition:  If there exists  $X$ such that $XA\ \equiv \ -(1+C) $,  
then  
$$ S T^X \ S  T^{-A }\ S  T^{-(D+BX)} \        \equiv  
   \pmatrix{ A  & B \cr C & D}                                 $$
Proposition:  If there exists  $X$ such that $XA \ \equiv  1-C $,  
then  
$$ S T^X \ S    T^{ A}\ S^{-1} \  T^{D+BX } \ \equiv  
   \pmatrix{ A  & B \cr C & D}                                 $$

Examples: 
      
$$    \pmatrix{ 3 & 4 \cr 2 & 3}
 \ =\  T^2 \ S    T^2 \ S    T^2  \ \ \ \   \hbox{ in } Z      $$ 
 
$$   \pmatrix{ 5 & -8 \cr 2 & -3} \ = 
   \  T^2 \ S    T^{-2} \ S^{-1} T^{-2} \ \ \ \  \hbox{ in } Z $$
\vskip .5cm 
{\bf If $(c,N)=1 $ } , we have both solutions 
$(U_+  ,V_+  )\  \equiv \ (\ (A+ 1)C^{-1}\ ,   (D+ 1)C^{-1} \ )   $ ,
         
and  $(U_-  ,V_-  )\  \equiv \ (\ (A- 1)C^{-1}\ ,(D- 1)C^{-1} \ ) $,
 where $ C^{-1} $ is the inverse of C mod N:       
$$  \pmatrix{A  & B \cr C & D} \equiv \  
T^{AC^{-1}}\  {\cal S}_C \   T^{DC^{-1}}\                      $$ 

$$ \hbox{where  }\ \  {\cal S}_C \ {\bf := } \  
       T^{C^{-1}  }\ S  T^{C}\ S T^{C^{-1} }\ \equiv           $$

$$  \equiv  T^{-C^{-1} }\ S  T^{-C}\ S^{-1}\  T^{-C^{-1} }\ 
\equiv  \pmatrix{0  & -C^{-1} \cr C & 0}  
\equiv \ ( \sigma_{ C^{-1}   } (S)\ , \ \hbox{see below } \ )  $$

\vskip .5cm    
These equalities reflect   some relations in the group 
 $SL2(Z/NZ) $  : 

$$ T^{2 C^{-1}}\ S  \  T^{C}\ S \ =  \ S\  T^{-C} \ S^{-1}  \  T^{-2 C^{-1}}$$

\vskip .5cm 

$$\hbox{Since   }  \ \  S\ \pmatrix{ A  & B \cr C & D}    \ S^{-1} 
   = \pmatrix{ D & -C \cr -B & A} \ ,   $$

only when the four residues $A,B,C,D$ are non invertible mod $N$, could we have 
 complicated expressions for  elements of   $SL2(Z/NZ) $. This could  happen  
 only when $N$ is not a prime power.


\vskip .5cm 

{\bf Results from Physical Approaches }

In the sequel, we will use an abelian  group of automorphisms (which we call  the 
$\sigma_L $'s ) of the group $SL2(Z/NZ)$ defined by:

$$    \sigma_{ L} \left(  \pmatrix{A  & B \cr C & D} \right)\ :\equiv  
   \pmatrix{A  & BL \cr CL^{-1} & D}                            $$
Obviously, this group of automorphisms is isomorphic to 
the multiplicative group of invertible residues mod N: 
\vskip .5cm 
Furthermore these  morphisms satisfy:
$$ \sigma_{ K}(S)\  \sigma_{ L}(S)\equiv  \sigma_{ K/L}(S)\ S
   \equiv \ S\      \sigma_{L/ K}(S)\equiv  \pmatrix{-K/L  & 0\cr 0 & -L/K} $$

{\bf Theorems}( de Boer, Goeree, A.C., Gannon, Lascoux, Bantay): 
             
 Any Rational Conformal Field Theory defines a representation  $\rho $ of 
 $SL2(Z) $ whose matrix elements $  \rho( M )_{pp' }   $ are in a 
 cyclotomic  field of N-th roots of unity, 
 $ \rho( T^N ) $  is the identity,  $\Gamma (N)$ lies in its kernel  , and for any 
matrix M of $SL2(Z/NZ) $, the above morphisms go into the cyclotomic characters: 
 $$ \sigma_{ L}(  \rho( M )_{pp' }  )\ =
                \ \rho (  \sigma_{ L}(M) )_{pp' }     $$

$$ \hbox{where on the l.h.s }\   \sigma_{ L}(\xi_N ) = \xi_N^L  $$ 

In previous texts,  explicit computations were 
given for the example of $sl(2) $ 
affine Lie algebra (so called Wess Zumino Witten models),   
Furthermore it was exposed how these theorems can lead to 
 very compact formulae for the representation on Virasoro 
characters.

\vskip 4.cm

{\bf Relations}

\vskip 1.cm 
In this paragraph we give examples of constructions of 
quotient groups as anounced  above:

We define the relations  ${\cal R}_N $ to be: 
$$ {\cal R}_N \  :  S^4=T^{N}=1\ ,\ (ST)^3 \ =S^2  $$
The group generated by S, T  and these relations, is for $N\geq 6$, 
associated to a tesselation by 
 triangles. The question of which relations can be added 
in order to obtain a quotient group which is finite,
 or a triangulated  surface of finite genus ( and finite area) naturally arises.
 Note that for 
some authors a Riemann surface is compact and 
 this is not the case as long as we keep the hyperbolic 
 metric structure for which the punctures are at ( logarithmically ) infinite distance.

Other questions are: 
 what are the relations which lead to the group $ SL2(Z/NZ) $ ?. 
At given N, what are the relations which give a surface of minimal genus? 
\vskip 1.cm 
Here we address only the first question for which we need the following: 
         
\vskip .5cm     

Lemma: 
when $ BC \equiv \ -2 $, 
$$   ST^{C}\  S T^{-B}\ \equiv 
      T^{B} \ S T^{-C} \ S^{-1}  \ \equiv   \pmatrix{ -1 & B\cr C & 1}  $$ 
when $BC \equiv \ 0 $, 
$$   ST^{C}\  S T^{-B}\  \equiv \  
     T^{-B}\ S T^{C} \ S\      \equiv \pmatrix{ -1 & B\cr C & -1} $$ 
$$ ST^{-C}\  S^{-1}\  T^{B}\  \equiv \  
    T^{B}\ S T^{-C} \ S^{-1}\  \equiv \pmatrix{ 1 & B \cr C & 1 } $$

{\bf Proposition: }   
$$ \  S T^U \ S  T^{- A}\ S \  T^{V } \  \equiv  \   
     T^X \ S    T^{- D}\ S \  T^{Y }\ S \  \  \  \hbox{If and only if }\     $$
$$  A  \equiv  \ -(X+BY)\ ,  \ D \equiv \ -(V+BU)\ ,\ UX  \equiv \ VY   $$
$$ \hbox{ where necessarily }\ B \equiv \   1+XD \        \equiv \ 1+AV $$

{\bf Proposition: let p be a prime}   
$$ \hbox{If }\  N = p^n >2 \ , \   SL2(Z/NZ)  \hbox{is  presented with extra relations:}  
  $$
$$ H_A \ := T^A   S T^{1/A} \ ST^A  \  S^{-1} \
 \hbox{for invertible} \ A's  \ , H_A\ H_B =H_{AB} $$
$$ H_A\ T= T^{A^2} \ H_A \  ,\  H_A \ S \ = S^{-1} \ H_{1/A} $$

Proof: define for each $(C,D) $ such that there exist $A,B$,  $AD-BC\equiv 1$, 
 a word  $X_{(C,D)} $ in S and T   which corresponds to both an  SL2 matrix and 
an hyperbolic triangle.  We can enumerate elements of the group mod N 
as words of the form  $ T^x \ X_{(C,D)} $.

  Then one proves that  $ SL2(Z/NZ)/\pm 1  $ is generated 
by relations  given above (with  $S^2 =1 $  ) by a careful study  of  glueing 
 formulae  at the boundary  of the  connected triangulated domain.
This is achieved by checking by use of the above relations that for each 
 $\ X_{(C,D)} $ , $ \ X_{(C,D)}\ T^{\pm 1} $  
and  $\ X_{(C,D)}\ S  $ are of the form  $T^L \ X_{(C',D')} $ 
for some $L,C',D'\ \in Z/NZ $ . That the relations do not 
give a smaller quotient comes from the fact easily checked 
that the matrix group $SL2(Z/NZ) $ explicitly do satisfy 
these relations. 
  
\vskip .3cm 
  
Since it deserves some time, let us give some explicit steps 
in a constructive  proof of the above presentation:  
  
$X_{(0,1) } \ = \    $ identity.  
      
$X_{(1,D) } \ = S T^D $

For $(c,N)=1$, $ 2\leq c\leq {N-1 \over 2} \  $ and for any D: 
$X_{(C,D) } \ = S T^C \ ST^{(D+1)/C} $. Note that for more general N one could also take 
 
$X_{(C,D) } \ = S T^C \ ST^X \ \equiv    \pmatrix{-1  & -X\cr C & D}  $
 whenever exists $X$ such that $CX\equiv D+1  $ ; note that this includes the cases   
 $X_{(C,-1) } \ = S T^C \ S   $.

For c not coprime with p (thus with N ) , $0\leq c \leq  {N\over 2} $ and 
$(d,N)=1 $ , $ 2\leq d\leq   {N-1 \over 2} $ we take 
$X_{(C,D) } \ = ST^d \ ST^{(1-c)/d}\ S^{-1}   $.
       
With the same c and $(d,N)=1 $ ,    $ -\ {N\over 2} < d\leq -2 $    
$$ X_{(C,D) } \ = ST^{-D} \ ST^{-(1+C)/D}\ S  
  \equiv        \pmatrix{(1+C)/D  & 1\cr C & D}   $$.
  
Finally for C non invertible, $2\leq c < N/2 $:
$X_{(C,1)} = ST^{-C}\ S^{-1}=   \pmatrix{1  & 0\cr C & 1}   $ 


This presentation by generators and relations being established, 
a few remarks are   useful:  
  
 A dual point of view which is also useful 
is to construct the surface by glueing  N-gons which are collections 
of triangles labelled by words $ Y_{(A,C)} \ T^z $. Then the centers of the 
corresponding N-gon can be seen as having coordinate 
$\tau = {a\over c }$ on the real axis ( boundary of the upper half plane).  
 A more rigourous formulation is of course to identify the center of the 
N-gon to
the orbit of $   {a\over c }  $ under homographic $\Gamma(N) $ tranformations. 

\vskip .3cm 

  Of course, the above
  relations in terms of the $H_A \ 's$ 
are redondant, it suffices to have them for generators of this abelian group
 (Cartan torus) isomorphic  to the group of invertible residues mod N. Even 
one can find in literature various relations, which we already collected in 
a previous electronic text with T. Gannon (arXiv/math.QA/9909080 ).
  According to various authors 
their relations do in fact  
 imply the above relations. (Such implication may come from interesting 
constructions and from the congruence subgroup property). 

 We give below  for small values of N, such 
  simple and compact looking presentations.

Note the genus of the Riemann surface is 
$$ \hbox{if }\  N= p^n \ , \  
   g=\ 1\ +\ {p^{3n}- p^{3n-2}-6 p^{2n} +6p^{2n-2} \over 24 } $$

The above construction allows us to enumerate explicitly elements of  
 $ SL_2(Z/NZ ) $, when N is a prime power. Of course an explicit use 
of the chinese remainder theorem gives us  a description of the general 
group $ SL_2(Z/NZ ) $ as direct product of its primary factors.

Let us give explicitly formulae for the generators in terms of 
decomposition of $1$ (mod $N$ ) into orthogonal sum of 
idempotents: This is textbook result for commutative semisimple algebras, 
  called in french ``alg\`ebres r\'eduites'' by Bourbaki:  

$$ 1\equiv \Sigma_{p, \ p|N}  \  c_p   \ \hbox{ mod } N 
   \ \ \  \ \ \ \   T_p :=  T^{c_p }                    $$
$$ S_p \  :=  S^2 \ (ST^{1-c_p } )^3 \ =  S^2 \ (T^{1-c_p }\ S )^3 $$

\vskip .5cm 

But a decomposition into primary factors gives a much too complicated 
 description  of elements of $ SL_2(Z/NZ ) $ as  words in $S$ and $T$. 
There is a much smarter approach!: 
 
 \vskip .5cm                 
 {\bf Proposition: }   
       
Any  element of $ SL_2(Z/NZ ) $ can be written with at most four 
powers of $T$, i. e. as a word like: 
 
$$    T^{x_3  }\  S  T^{x_2  }\  S  T^{x_1  }\  S  T^{x_0  }    $$ 

{\bf Proof: }  We start with the following {\bf lemma}: 
       
Let a,b,c,d, N be five integers satisfying $ad-bc=1$\ ,\  $N>0 $ .
 Then there exists an integer 
 m such that $d' \ := d-mc $ is coprime to $N$.  
Then denote ${\bf D' }$ its residue mod $N$ , and $D'^{-1} $ its inverse.   
              
 Dirichlet has even proven that one could find  an infinity of 
values of m such that $d' $  is a 
prime number   
not dividing N; but here the requirement is much weaker,
 so that one can find a convenient m again 
 with help of the chinese
 remainders,   because that means there exist $u,v$  such that 
  $u(d-mc)\ -vN \ =1 $ which is equivalent to the existence of a residue 
  $M_p $   
modulo each $p^{\nu } $ factor of N such that the residue 
  $ D_p - \ M_p C_p $ is invertible mod     $p^{\nu } $. 
If  $D_p $ is invertible, $M_p =0 $ works, and if  $D_p $
 is not invertible  , $C_p $ is and therefore any $M_p $
  which is not divisible by $p$ will do the job.

{\bf Then } we have ($b'=b-ma$ , $d'= d-mc$ ):  

$$     \pmatrix{a  & b\cr c &d } = \pmatrix{ a & b-ma\cr c & d-mc } \ 
       \pmatrix{1  & m\cr 0 &1 } $$
$$ \hbox{Thus }\ \    \pmatrix{A  & B\cr C &D } 
\equiv  T^{(B'-1)/D'}  \ S  T^{-D'} \ S  T^{-(1+C'\  )/D'} \ ST^{M}
  \ \   \ \hbox { mod } \  N   $$

\vskip .5cm 
Physicists' intuition would find it natural since $SL2(R) $ 
is a three dimensional manifold, 
and indeed here we have succeeded in decomposing any group element 
into an expression with four parameters  
so there is some kind of one-parameter degree of arbitrariness.  Nevertheless this 
is too  naive a picture because $SL2(Z) $ is really exceptional: one needs unbounded 
words to express any matrix, the continued fraction of any rational $a/c$ can be of 
any arbitrary length. This is even exceptional, compared to $SL3(Z) $.   
  
\vskip .5cm 
 
As a {\bf conclusion} we could say that the groups $SL2(Z/NZ) $ appear in fact simpler 
than what could be anticipated from  $SL2(Z)$. This allowed us to improve slightly 
upon prior works, bringing  our little stone to the building. 
\vskip .5cm 
    
 Nevertheless there is a fascinating 
interplay between geometry and number theory, as usual due to the 
apparently chaotic  occurence of primes and congruences when one decomposes 
an integer, N , into its prime factors.
If we were considering strings, membranes or black holes we could claim that  
 Conformal Field Theories bring
more pieces into some   cosmic hyperbolic  puzzle! We prefer to let the reader appreciate 
the intrinsic beauty of mathematics and rigorous physics.

\vskip 1.cm

{\bf Examples:} 
    
We finally give explicit presentations for small values of $N$: 
 $N=5 $ is the famous F. Klein's icosaedron (or dodecaedron). For
 $ N=6$ we have a torus, which can be equivalently defined  by the two 
presentations below.    Another very interesting approach from a geometric 
point of view  is to 
identify fundamental domains as done in places like Bonn by Kulkarni( see refs ). 
Examples of quotients from conformal theories are detailed in previous 
texts by A. C. 

 $$ SL_2(Z/5Z )\  = <S,T\ | {\cal R}_5  \ >              $$
$$ SL_2(Z/6Z )\  = <S,T\ | S^4=T^{6}=1,\ (ST)^3 \ =S^2,\ ST^{2}\ ST^{-2}\ 
   = T^{2} \ S T^{-2} \ S  > $$ 
$$ SL_2(Z/6Z )\  = <S,T\ | S^4=T^{6}=1,\ (ST)^3 \ =S^2,\ ST^{3}\ ST^{2}\ 
   = T^{2} \ S T^{3} \ S  > $$ 
                           
$$ SL_2(Z/8Z )\  = <S,T\ | S^4=T^{8}=1,\ (ST)^3 \ =S^2,\ ST^{2}\ ST^{4}\ 
   = T^{4} \ S T^{2} \ S  > $$ 

 $$ SL_2(Z/9Z )\  = <S,T\ | S^4=T^{9}=1,\ (ST)^3 \ =S^2,\   
   ST^{3}\ ST^{-2}\ S^{-1}\    = T^{-4} \   ST^{-2}\ S T^{-2}           $$
$$     \ ,\   (S T^{3 }\  )^2\ =\  (T^{3 } \ S )^2 \  , 
       \ ST^{4}\ ST^{-4}\      = T^{4} \ S T^{-4}\ S^{-1}  \ ,          $$
$$ ST^{-2}\ ST^{4} \ ST^{-2}\  = T^{-4}\  ST^{2 }\ S T^{-4} \ S^{-1}\ , $$
$$ ST^{2 }\ ST^{-2}\ ST^{4 }\  = T^{2 }\  ST^{-4}\ S T^{3 } \ S^{-1} >  $$ 

 $$ SL_2(Z/10Z )\  = <S,T\ | S^4=T^{10}=1,\ (ST)^3 \ =S^2,\ ST^{2}\ ST^{5}\ 
        = T^{5} \ S T^{2} \ S   $$ 
$$ ,\   ST^{3}\ ST^{4}\ 
       = T^{-4} \ S^{-1}\  T^{-3} \ S \ ,\  
 ST^{3}\ ST^{-3}\ ST    = T^{-1} \ S T^{3} \ ST^{-3}\ S  $$ 
$$ \ ,\     ST^{4}\ ST^{5}\  = T^{5} \ S T^{4} \ S\   >\ \hbox{ genus }13  $$


\vskip .5cm


{\bf References and Acknowledgements }
A.  C. thanks   J. Lascoux, H. Behr , T. Gannon , P. Ruelle, J. Wolfart, 
J-C. Schlage-Puchta, J. Froehlich, V. Pasquier, D. Zagier,
 for help and many conversations,     
M. Burger,   M. Kraemer for very kind hospitality  in  FIM ETHZ.

P. Bantay, ``The kernel of the modular representation and the Galois action in 
RCFT'', Communications in Math. Phys. 233 (2003) 423-438, and further texts.

A. Coste,  ``On Rational Conformal Field Theories:
 Explicit Modular Formulae''  math-phys/0405,    
Letters in Mathematical Physics, Dijon (2005) vol72, 1-15.

A. Coste  ``Investigations sur les caracteres de Kac Moody et 
  certains quotients de SL2(Z)'' (in french) 
   preprint  IHES/P/1997/78,  unpublished.

A. Coste, T. Gannon, Physics Letters B 323 (1994) 316.

J.M. Drouffe, C. Itzykson, ``Statistical Theory of Fields (10 chapters)'' , 
Ediscience, and references therein.    

J. de Boer, J. Goeree, Communications in Math. Phys. 139, (1991) 267.

E. Hecke , Elementary number theory, (translated from German 
and reprinted) Springer GTM.

Ravi S. Kulkarni, 
``An arithmetic geometric study of the subgroups of the modular group'', 
American Journal of Math. 113 (1991), 1053-1133.

H. Rademacher, Analytic Number Theory, Springer.

\end